\documentclass[a4paper,11pt]{article}
\pdfoutput=1
\usepackage{jcappub}

\usepackage[english]{babel}

\usepackage{graphicx}
\usepackage{amsmath,amssymb}
\usepackage{braket}
\usepackage{subfigure}
\usepackage{float}
\usepackage[dvipsnames]{xcolor}
\usepackage{mathrsfs}
\usepackage{comment}
\usepackage{multirow}
\usepackage{soul}
\usepackage{mathtools}
\usepackage{sidecap} 
\graphicspath{{Figures/}}
\usepackage[normalem]{ulem}
\usepackage{array}
\usepackage{mathtools}

\title{From Evidence to Evident:  Decisive Cosmological Evidence for the Normal Neutrino Mass Hierarchy}

\author[a,b]{Raul Jimenez,}
\affiliation[a]{ICC, University of Barcelona, Mart\' i i Franqu\` es, 1, E08028
Barcelona, Spain}
\affiliation[b]{ICREA, Pg. Lluis Companys 23, Barcelona, 08010, Spain.} 

\author[c]{Carlos Pe\~na Garay,}
\affiliation[c]{Laboratorio Subterr\'aneo de Canfranc, 22880 - Canfranc-Estaci\'on , Huesca, Spain.}
\author[a]{Fergus Simpson,}

\author[a,b]{Licia Verde}

\emailAdd{raul.jimenez@icc.ub.edu; cpenya@lsc-canfranc.es; fergus2@gmail.com; liciaverde@icc.ub.edu}

\abstract{
Cosmological data have now reached the precision required to turn the neutrino mass ordering
from a weak Bayesian preference into a decisive model-selection test. We compute the evidence
for the Normal and Inverted Hierarchies by combining the DESI DR2 clustering analysis with NuFIT  oscillation data. In the baseline $\Lambda$CDM model, DESI DR2
and Planck CamSpec give an upper limit for  the sum of neutrino masses  $\Sigma m_\nu<0.0642\,{\rm eV}$ at 95\% confidence, close to the
normal-ordering floor, $\Sigma m_\nu^{\rm NH}\simeq0.059\,{\rm eV}$, but well below the
inverted-ordering minimum, $\Sigma m_\nu^{\rm IH}\simeq0.099\,{\rm eV}$. This constraint places the
entire inverted hierarchy in the tail of the cosmological likelihood. The resulting Bayes factor,
$K=P(D|{\rm NH})/P(D|{\rm IH})$, exceeds $460$ even for a (conservative) reference prior,
and remains strong, $K>40$, in extension of the baseline cosmological model. 
We show that this conclusion is
robust to the choice between a reference prior and a physically motivated, logarithmic 
hierarchical prior, marking the transition from {\it prior-sensitive evidence} to {\it likelihood-dominated
exclusion} of the inverted hierarchy within standard cosmology. By embedding these two priors in
the full two-dimensional design space of measure (logarithmic versus linear in mass) and structure
(hierarchical versus non-hierarchical), we show that all four prior constructions yield decisive
evidence under DESI DR2, and that the residual prior dependence is governed almost entirely by the
choice of measure---a factor $\sim\!10$ in the Bayes factor---rather than by the hierarchical
assumption. At the prior-family level, the same evidence calculation consistently favors the SJPV prior predictive over HS by a Bayes factor of over $4,700$ across each of the matched-support variations tested. The favored normal ordering
pushes the effective Majorana mass to the few-meV regime, with median
$m_{\beta\beta}=3.28\,{\rm meV}$ and 95\% credible interval
$0.95<m_{\beta\beta}<11.55\,{\rm meV}$, below the canonical inverted-ordering
target for upcoming neutrinoless double-beta decay experiments. 
}

\begin{document}
\maketitle

\section{Introduction}
\label{sec:intro}

Neutrino oscillations have established that at least two neutrino mass eigenstates are non-zero, but they do not determine the absolute mass scale, the Majorana or Dirac nature of neutrinos, nor the ordering of the mass eigenstate with the largest electron neutrino component~\citep{Fukuda:1998mi,Ahmad:2002jz}. 

The latter question is usually phrased as the problem of mass ordering. In the normal hierarchy (NH), the single state most separated by the atmospheric splitting is the heaviest one, so that $m_1<m_2\ll m_3$ in the hierarchical limit; in the inverted hierarchy (IH), the single state is the lightest one, $m_3\ll m_1<m_2$. In addition, oscillation data remain insensitive to the common offset in the three masses. This leaves unconstrained a one-dimensional absolute-mass direction, conveniently parameterized either by the lightest mass or by the sum of the three masses, 
\begin{equation}
\Sigma m_\nu \equiv m_1+m_2+m_3 .
\end{equation}
For the current best-fit oscillation parameters~\cite{Esteban_2026,nufit-6.1}, the two possible mass orderings imply different irreducible lower limits: $\Sigma m_\nu\simeq 0.059\,\mathrm{eV}$ for NH and $\Sigma m_\nu\simeq 0.099\,\mathrm{eV}$ for IH. This separation is small in absolute terms but large enough that precision cosmology can, in principle, 
answer a question of flavor physics from measurements of cosmology clustering.

Cosmology is sensitive to neutrino mass because massive neutrinos free-stream while being relativistic and cluster inefficiently after becoming non-relativistic. Their effect is imprinted on the cosmic microwave background, on the late-time expansion history and, most directly, on the suppression of the matter power spectrum below the free-streaming scale~\citep{Lesgourgues:2006nd}. The observable most robustly constrained by cosmology is the total mass $\Sigma m_\nu$, rather than individual eigenvalues.

This makes cosmology complementary to oscillation experiments: oscillations define the allowed NH and IH hypersurfaces in mass space, while cosmology restricts the projection along $\Sigma m_\nu$~\citep{Jimenez_2010}. As the upper limit on $\Sigma m_\nu$ approaches and then passes below the minimum value allowed by IH, the inverted spectrum is compressed against a physical boundary and reduces likelihood support.

The neutrino mass hierarchy is a boundary condition for theories of lepton flavour and for mechanisms for generating neutrino mass. Models based on seesaw dynamics, flavour symmetries, radiative mass generation, or higher-dimensional operators often encode assumptions about whether the three families arise from a common structure or from more fragmented physics. In this sense, the ordering problem is a particularly clean arena in which empirical model selection and theoretical prior information meet.

The neutrino mass hierarchy problem is naturally formulated as a Bayesian comparison between two composite hypotheses, each with its own parameter volume and evidence integral. This Bayesian formulation has a substantial history, and the role of the prior has been central since the beginning. Early combinations of cosmology and oscillation data produced only weak or moderate ordering preferences, because the cosmological upper limits were far above both oscillation floors~\citep{Hannestad:2016fog,Gerbino:2016sgw}. The situation changed qualitatively with the analysis of Ref.~\cite{Simpson:2017} (hereafter SJPV), who argued that a minimally informative prior on the individual neutrino masses, together with cosmological limits on $\Sigma m_\nu$, implied strong Bayesian evidence for NH. SJPV is particularly instructive here
not only for its numerical odds but also because it made explicit a geometric effect: when the three masses are treated as the primitive variables, the oscillation constraints carve two thin allowed filaments through mass space, and the IH filament occupies a different prior volume than the NH filament once the cosmological bound pushes the allowed region toward the minimum-mass regime. The result stimulated a useful debate about whether the apparent preference was data-driven or prior-driven~\citep{Schwetz:2017fey,Heavens:2018adv,Gariazzo:2022evs}. That debate is directly relevant to the present paper, because it clarified that the phrase ``evidence for the hierarchy'' is incomplete unless one specifies the measure on mass space, the treatment of the oscillations and the cosmological likelihoods.

The follow-up work by~\cite{Jimenez:2022dkn} sharpened this point and is the immediate methodological precursor of the present analysis. It showed that once the cosmological constraint on $\Sigma m_\nu$, further tightened by new data, is combined with contemporary laboratory information, the normal hierarchy is favoured at decisive odds over a wide range of prior choices. Crucially,  ~\cite{Jimenez:2022dkn} did not rely on a single subjective prior prescription. Instead, they compared two deliberately different extremes: a physically motivated hierarchical prior in the spirit of SJPV, and an objective reference prior designed to minimise information not supplied by the measured oscillation quantities~\cite{Heavens:2018adv}. The conclusion was that the numerical Bayes factor is prior-dependent, as it must be, but that the qualitative inference remained stable: the odds for NH exceeded $140:1$ across the prior choices explored. The present paper extends this programme to the DESI Data Release 2 era, where the cosmological likelihood has moved from a regime where the IH only occupies a smaller prior volume than the NH to one in which the IH lower bound lies in direct tension with the preferred cosmological mass range.

The new ingredient is the precision of the DESI DR2 baryon acoustic oscillation measurements in combination with the CMB information. In the baseline $\Lambda$CDM analysis used here, DESI DR2 plus Planck CamSpec produces a marginalized limit $\Sigma m_\nu<0.0642\,\mathrm{eV}$ at $95\%$ confidence~\citep{DESI:2025neutrino,DESI:2025bao}. 
Taken at face value, this upper bound
lies only slightly above the NH floor and well below the IH floor. The resulting hierarchy test is therefore qualitatively different from earlier applications (but see~\cite{Heavens:2018adv} for a dissenting/contrasting view): the evidence is no longer controlled only by how much prior volume remains near the IH minimum, but also by the fact that the cosmological posterior itself is centered in a region difficult to accommodate with IH.

This finding has immediate experimental consequences.
If neutrinos are Majorana fermions, the rate of neutrino-less double-beta decay depends on the hierarchy.

The IH generically predicts an experimentally accessible band, whereas NH permits stronger cancellations and values in the few-meV range \citep{DellOro:2016tmg}. Therefore, a cosmological exclusion of IH would imply requirements not within reach of the largest sensitivity double beta decay experiments KamLAND2-Zen~\citep{KamLAND2Zen}, LEGEND1000\citep{LEGEND:2021bnm} and CUPID\citep{CUPID:2019imh} to fully explore all neutrino mas scenarios allowed by current data. Conversely, an observed signal in the conventional IH band would become increasingly difficult to reconcile with the standard and widely accepted cosmological inference.

In this context, DESI DR2 has highlighted a broader issue in cosmological neutrino inference: depending on dataset combinations and model assumptions, the inferred effective neutrino mass can be driven close to, or formally below, the physical lower bound implied by oscillations. This makes it essential to distinguish three questions that are sometimes conflated: 1) how strongly (and how robustly) cosmology constrains $\Sigma m_\nu$ within a chosen cosmological model; 2) how strongly the data prefer NH over IH once oscillation information is imposed; and 3) how robust that preference is to the measure adopted on the neutrino mass parameters.

Here we will not be concerned on the robustness of the DESI neutrino mass results and we will take the published  constraints face value. The results presented here generalize in a straightforward way to possible future corrections  on the cosmological neutrino mass bound. 
For this reason, the present analysis is structured around robustness to prior assumptions.

The paper is organized as follows. We first construct an analytic representation of the DESI DR2 cosmological likelihood for $\Sigma m_\nu$ and combine it with the latest NuFIT global analysis of neutrino oscillation data. We then compute the evidence ratio $K=P(D|\mathrm{NH})/P(D|\mathrm{IH})$ under the SJPV and HS priors, reconstructing the historical evolution of the Bayes factor from the early cosmological limits to the DESI DR2 regime. 
To make the role of the prior fully explicit, we embed these two prescriptions in the complete two-dimensional design space spanned by the choice of \emph{measure} on the masses (logarithmic versus linear) and the choice of \emph{structure} (hierarchical versus non-hierarchical), constructing the two previously unexplored corners and showing that the residual prior dependence is controlled by the measure rather than by the hierarchical assumption. 
Next, we derive the posterior distributions of the individual masses and propagate the hierarchy inference to the effective Majorana mass $m_{\beta\beta}$. Finally, we discuss the dependence on the assumptions of the cosmological model, including extensions beyond the flat $\Lambda$CDM.

\section{Methods}
\label{sec:methods}

\subsection{Cosmological and neutrino oscillation data and their likelihoods}
Our analysis combines the latest cosmological constraints on the sum of neutrino masses, $\Sigma m_\nu$, with data from neutrino oscillation experiments.

The foundation of the evidence calculation is a faithful analytical representation of
the cosmological likelihood for the total neutrino mass,
$P(D_{\rm cosmo}\,|\,\Sigma m_\nu)$, derived from the publicly released DESI DR2
Markov Chain Monte Carlo (MCMC) chains combining DESI BAO with the Planck CamSpec
CMB likelihood within the baseline flat $\Lambda$CDM cosmological model
\citep{DESI:2025neutrino,DESI:2025bao,Planck:2018vyg,Rosenberg:2022sdy}.
The marginalized one-dimensional posterior on $\Sigma m_\nu$ implies an upper limit
\begin{equation}
\Sigma m_\nu < 0.0642~\mathrm{eV} \qquad (95\%~\mathrm{C.L.}),
\label{eq:DESI_DR2_limit}
\end{equation}
a value that lies within a few~meV of the lower bound allowed by NH and well below
the IH floor. This proximity to the physical boundary is the central characteristic that
governs every subsequent inference and requires a careful analytical treatment
that respects the prior $\Sigma m_\nu \ge 0$ without introducing artifacts associated
with truncation in the bulk of the posterior~\citep{Feldman:1997qc}.

To enable a continuous, smooth evaluation of the evidence integrals while preserving
the relevant features of the underlying MCMC distribution, we model the marginalized
posterior as a truncated Gaussian density,
\begin{equation}
P_{\rm trunc}(\Sigma m_\nu\,|\,\mu_0,\sigma)
= \frac{1}{\mathcal{N}(\mu_0,\sigma)}\,
\frac{1}{\sqrt{2\pi}\,\sigma}\,
\exp\!\left[-\frac{(\Sigma m_\nu-\mu_0)^2}{2\sigma^2}\right]\,
\Theta(\Sigma m_\nu),
\label{eq:truncgauss}
\end{equation}
where $\Theta$ is the Heaviside step function and $\mathcal{N}(\mu_0,\sigma)$ is the
normalization in the physical region $\Sigma m_\nu\ge0$.

We reconstruct the chain posterior with a kernel density estimate (KDE), and we also perform a Feldman--Cousins-style profile
likelihood  obtained by minimizing $\chi^2$ along orthogonal
directions in the joint parameter space which optimal parameters are
determined by simultaneously matching the 95\% upper limit and the local curvature.
Both are described by a truncated gaussian with \begin{equation}
\mu_0 = -0.036~\mathrm{eV}, \qquad
\sigma = 0.043~\mathrm{eV}.
\label{eq:truncpars}
\end{equation}

\begin{figure}[h!]
    \centering
    \includegraphics[width=0.6 \linewidth]{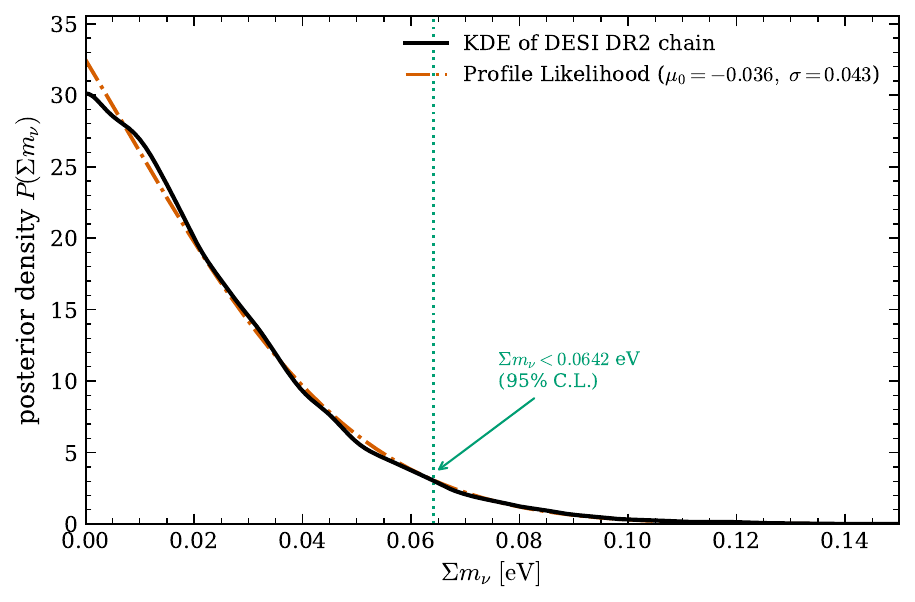}
    \caption{Marginalized one-dimensional posterior probability density for the
    sum of neutrino masses $\Sigma m_\nu$, obtained from the DESI DR2 BAO and
    Planck CamSpec CMB likelihood combination within the baseline $\Lambda$CDM
    model. The KDE reconstruction of the MCMC chain (solid black) is compared to a Feldman--Cousins-style profile likelihood (dash-dotted orange; $\mu_0=-0.036~\mathrm{eV}$,
    $\sigma=0.043~\mathrm{eV}$). The vertical dotted line marks the
    95\% upper limit, $\Sigma m_\nu < 0.0642~\mathrm{eV}$. The 
    consistency of the two reconstructions across the physical region validates
    the analytical approximation for the likelihood used throughout the analysis.}
    \label{fig:likelihood_comparison}
\end{figure}

The slightly negative value of $\mu_0$ is purely a parametric reflection of the
fact that the cosmological likelihood, viewed as a function of $\Sigma m_\nu$
on the entire real line, is consistent with zero and indeed prefers values below the
physical boundary. 
Once the prior $\Sigma m_\nu \ge 0$ is imposed, the resulting
truncated distributions  are  shown in 
Figure~\ref{fig:likelihood_comparison}  across the relevant integration range.
We have also verified that adopting an
exponential or half-Gaussian functional form yields Bayes factors that differ
from the truncated Gaussian result by less than $\mathcal{O}(10\%)$, well below
the dynamic range of the inference reported below.

This shows that 
 the cosmological likelihood is well described by a near-Gaussian behavior
truncated at the physical boundary, and that the inference is not dominated by
non-Gaussian tails of the MCMC distribution. Frequentist and Bayesian summaries of
$\Sigma m_\nu$ from the same chains are known to differ  up to 
tens of percent level when the posterior abuts the physical boundary
\citep{Loureiro:2018pdz,Heavens:2018adv,DiValentino:2021hoh}: 
the close
correspondence shown in Fig.~\ref{fig:likelihood_comparison} demonstrates that
this difference is fully absorbed into the truncated-Gaussian
parametrization without distorting the evidence calculation.
 In what follows we therefore take this truncated Gaussian as the likelihood for $\sum_{m_\nu}$ for the evidence calculation. 

Within this framework, the recent DESI DR2 analysis has placed
the strongest cosmological bound on $\Sigma m_\nu$ to date, surpassing the
sensitivity of previous BOSS, eBOSS and Planck-only configurations
\citep{eBOSS:2020yzd,Planck:2018vyg,DiValentino:2021hoh}, and bringing the
neutrino mass constraint into the immediate vicinity of the minimal-NH floor.
The likelihood as described by Eqs.~(\ref{eq:truncgauss})--(\ref{eq:truncpars}) is
therefore a faithful proxy for the up-to-date cosmological information,
formulated in a manner that allows reproducible computation of the Bayesian
evidence for each hierarchy.

To test the sensitivity of our results to the underlying cosmological model, we also perform the analysis using the MCMC chains from a $w_0w_a$CDM model, which allows for a time-varying dark energy equation of state. In this case we obtain:
\begin{equation}
\mu_0 = 0.032~\mathrm{eV}, \qquad
\sigma_{w_0w_a}=0.051~\mathrm{eV}
    \end{equation}

\begin{figure}
\centering
\includegraphics[width=0.79\linewidth]{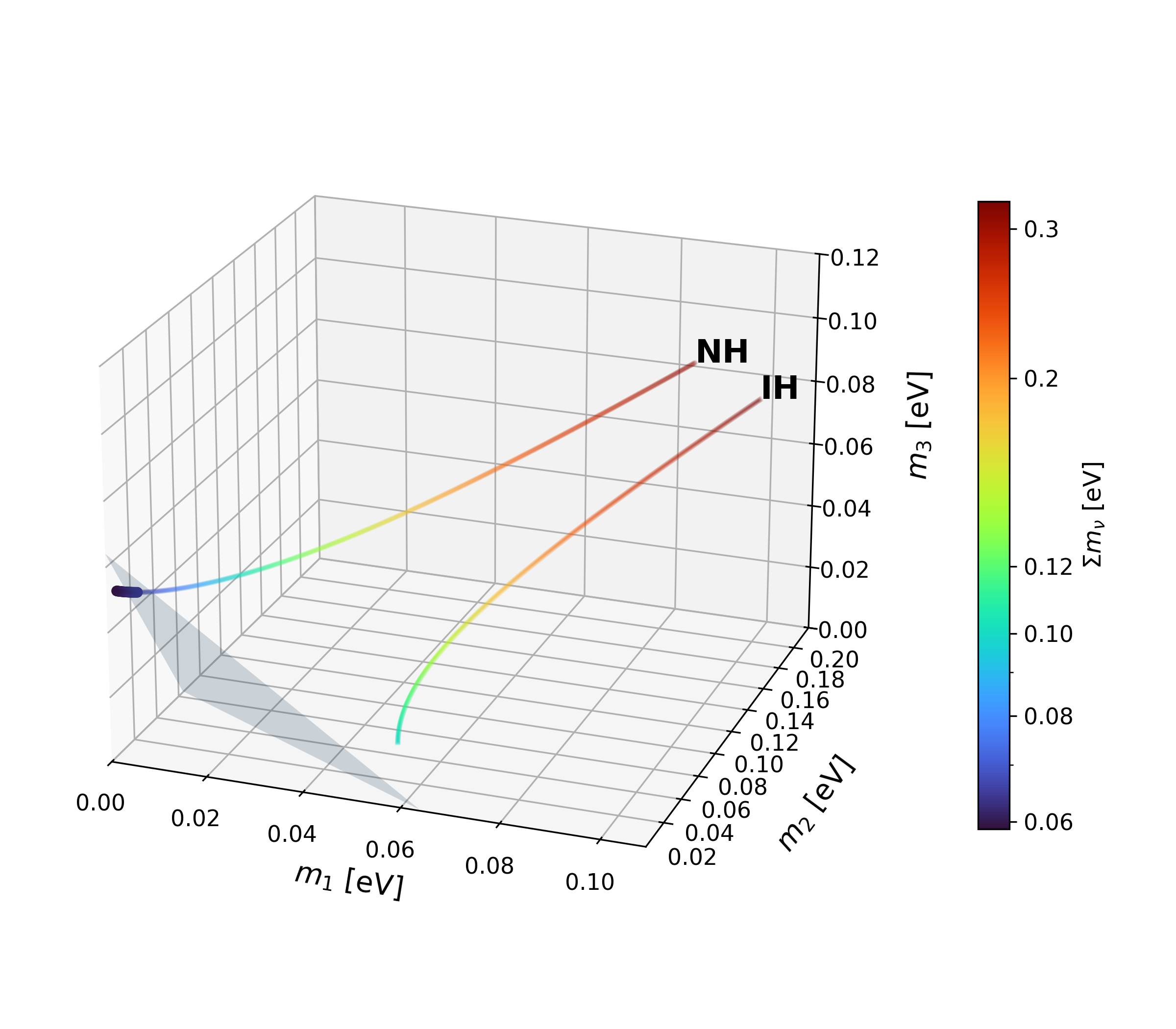}
\label{fig:filaments}
\caption{Oscillation constraints shown as filaments in neutrino-mass space, together with a   cosmological mass-sum plane at $\sum_{m_\nu}=0.0642$ eV.}
\label{fig:neutrino-filaments}
\end{figure}

The constraints on the neutrino mass-squared differences ($\Delta m^2_{21}$ and $\Delta m^2_{3\ell}$) and the leptonic mixing angles are taken from the latest NuFIT global analysis~\citep{Esteban_2026,nufit-6.1} based on oscillation data available at the end of 2025, which included the first results from the JUNO reactor neutrino experiment~\citep{JUNO_2025}. This information is incorporated into our framework as a likelihood component, $P(D_{\text{osc}}|\theta)$, where $\theta$ represents the fundamental neutrino mass parameters. The NuFIT analysis also provides a slight preference for NH with respect to IH, based on the global analysis of the available neutrino oscillation data from solar, reactor, atmospheric and accelerator neutrino experiments, corresponding to $\Delta\chi^2 = 9.4$, (the best fit parameters for the mass squared splittings in NH are $\Delta m^2_{21}=(7.53_{-0.10}^{+0.10})\times 10^{-5}~\text{eV$^2$}$, $\Delta m^2_{31}=(2.529_{-0.021}^{+0.021})\times 10^{-3}~\text{eV$^2$}$ and $\Delta m^2_{32}=(2.519_{-0.020}^{+0.017})\times 10^{-3}~\text{eV$^2$}$, while the best fit parameters for the mass squared splittings in IH are $\Delta m^2_{21}=(7.53_{-0.10}^{+0.10})\times 10^{-5}~\text{eV$^2$}$, $\Delta m^2_{31}=(-2.515_{-0.025}^{+0.031})\times 10^{-3}~\text{eV$^2$}$ and $\Delta m^2_{32}=(-2.484_{-0.026}^{+0.026})\times 10^{-3}~\text{eV$^2$}$), which is included in our final calculation of the Bayes factor. Figure \ref{fig:filaments} illustrates the NuFIT constraints in the volume defined by the neutrino masses $m_1, m_2, m_3$ and shows the correspondence to $\sum_{m_\nu}$.

\subsection{Bayesian model comparison framework}
The central goal of this work is to perform a Bayesian model comparison \citep{thorngren2025bayesianmodelcomparisonsignificance} between  NH and IH.
We compute Bayesian evidence, or marginal likelihood, for each hierarchy ($M \in \{\text{NH}, \text{IH}\}$), which is defined as the integral of the likelihood over the prior parameter space:
\begin{equation}P(D|M) = \int P(D|\theta, M) P(\theta|M) d\theta,\end{equation}
where $D$ represents the combined cosmological and oscillation data, and $\theta$ are the parameters that define the neutrino mass spectrum. The total likelihood is the product of the cosmological and oscillation likelihoods:
\[P(D|\theta, M) = P(D_{\text{cosmo}}|\Sigma m_\nu(\theta)) \times P(D_{\text{osc}}|\theta)\]
\citep{giusarma2012testingstandardnonstandardneutrino,barua2026cosmologicalconstraintsneutrinomasses}. The term $P(\theta|M)$ is the prior probability distribution of the parameters in the assumed hierarchy \citep{barua2026cosmologicalconstraintsneutrinomasses}.

The relative evidence for the two competing models is quantified by the Bayes factor $K$, defined as the ratio of their marginal likelihoods \citep{chib2016bayesfactorconsistency,mattei2020parsimonioustourbayesianmodel,kim2024deepbayesfactors}:
\begin{equation}K = \frac{P(D|\text{NH})}{P(D|\text{IH})}.\end{equation}
The value of $K$ is interpreted using the Jeffrey scale, where $K > 10$ indicates ``strong'' evidence and $K > 100$ indicates ``decisive'' evidence in favour of the Normal Hierarchy. \citep{dudbridge2024scaleinterpretationlikelihoodratios}

\subsection{Prior choices for neutrino masses}
The choice of prior, $P(\theta|M)$, is a critical component of any Bayesian model comparison \citep{bayarri2012criteriabayesianmodelchoice,llorente2022safeusepriordensities}. To ensure the robustness of our conclusions, we employ two distinct and well-motivated prior frameworks.

The first is the  physically-motivated hierarchical SJPV ~\cite{Simpson:2017} prior, also used in \cite{Jimenez:2022dkn}. 
This framework assumes that the three neutrino mass eigenstates, $m_1, m_2, m_3$, are exchangeable and drawn from a common underlying log-normal distribution.  Exchangeability encodes the 
pre-data symmetry that, absent a mechanism that singles out one eigenstate, the
prior over the unordered spectrum should not depend on arbitrary labels. However under an exchangeable prior on the primitive masses, the cosmological preference
for low \(\Sigma m_\nu\) translates into a geometric prior-volume penalty for the
IH filament.
The independent draws from a common underlying distribution reflects the physical hypothesis of a single, unified mechanism for mass generation. This structure inherently penalizes models that require fine-tuning of the mass parameters. The choice of lognormal distribution is motivated by the fact that before any information on the squared mass splitting, each of the neutrino masses has an uncertainty that spans many orders of magnitude. A uniform prior artificially favors the highest order of magnitude and may skew the interpretation of the results.  A logarithmic prior on the individual masses naturally reflects this scale uncertainty and is commonly regarded as the least-informative~\cite{clarke1994jeffreys}.

As a result, in this prior, IH which requires two nearly degenerate heavy masses ($m_1 \approx m_2 \gg m_3$), occupies a much smaller volume in the hyperparameter space compared to NH, resulting in a significant geometric volume penalty against IH \citep{Schwetz:2017fey}.

The second is an information-theoretic reference prior, hereafter referred to as the HS prior \cite{Heavens:2018adv}, constructed to be minimally informative by maximizing the influence of the data following the Bernardo-Berger framework. It is derived from the Fisher information matrix of the oscillation observables ($\Delta m^2_{21}$, $\Delta m^2_{3\ell}$). The resulting prior in the mass eigenstates ($m_L, m_M, m_H$ for the lightest, middle, and heaviest) is proportional to the Jacobian of the transformation from the observable basis, $P_{\text{HS}} \propto m_L m_M + m_L m_H + m_M m_H$. This prior does not assume exchangeability. Dropping exchangeability means that the prior no longer treats the three masses
as re-labellings of the same kind of object. Instead, at least one mass eigenstate
is assigned a different prior status from the others. As such, this prior does not impose a geometric penalty on the degenerate mass spectrum of the IH, providing a conservative baseline for our evidence calculation.

\subsection{Derived observables and evaluation}
We derive posterior probability distributions for several key physical quantities. The posteriors for the individual neutrino mass eigenstates ($m_1, m_2, m_3$) are computed for the 
selected case by combining the cosmological likelihood with the priors and oscillation constraints.

We also compute the posterior distribution for the effective Majorana mass, $m_{\beta\beta}$ \citep{benato2015effectivemajoranamassneutrinoless}, which governs the rate of neutrinoless double-beta decay \citep{chakraborty2025highlypredictiveneutrinomodel}. It is defined as:
\begin{equation}m_{\beta\beta} = \left| \sum_{i=1}^{3} U_{ei}^2 m_i \right|,\end{equation}
where $U_{ei}$ are elements of the Pontecorvo-Maki-Nakagawa-Sakata (PMNS) leptonic mixing matrix \citep{Pontecorvo:1957cp,Maki:1962mu}. We calculate the posterior for $m_{\beta\beta}$ using a Monte Carlo procedure. For each point in the posterior of the mass eigenstates, we sample the PMNS mixing angles and the Dirac CP phase from their distributions as given by the NuFIT covariance matrix, and we sample the two unknown Majorana CP phases uniformly over the interval $[0, 2\pi]$. This allows us to project the implications of our findings onto the sensitivity of current and future neutrinoless double-beta decay experiments.

\section{Results}
\label{sec:results}

This section presents the quantitative result of the Bayesian model comparison
between the Normal Hierarchy (NH) and the Inverted Hierarchy (IH).

First, to contextualize the impact of the DESI DR2 data, we perform a historical analysis (\S\ref{subsec:historical}) by re-computing the Bayes factor using a series of progressively tighter cosmological upper limits on $\Sigma m_\nu$ reported from 2002 to the present day. Complementing the findings and forecasts of \cite{Jimenez:2022dkn}, this demonstrates the evolution of statistical evidence as a function of the precision of cosmological data \citep{ong2026bayesianviewdesidr2}.
We then assess the role and stability of the prior
choice (\S\ref{subsec:prior_volume}), derive the posterior distributions of the
individual mass eigenstates (\S\ref{subsec:mass_posteriors}), and finally propagate
the hierarchy inference to the effective Majorana mass relevant for neutrinoless
double-beta decay (\S\ref{subsec:mbb}). A summary of the robustness of the result
under a $w_0w_a$CDM extension closes the section (\S\ref{subsec:w0wa}).

\subsection{Historical evolution of the hierarchy evidence}
\label{subsec:historical}

The strength of the present-day result is best appreciated by tracing the
historical evolution of the Bayes factor, since cosmological upper limits on
$\Sigma m_\nu$ have tightened over the last two decades. In spirit similar to Ref~\cite{Jimenez:2022dkn} we  re-evaluate the
evidence ratio
\begin{equation}
K \equiv \frac{P(D|\mathrm{NH})}{P(D|\mathrm{IH})}
\end{equation}
under a fixed methodological framework, identical priors, identical oscillation
inputs from NuFIT~\citep{Esteban_2026,nufit-6.1}, and
identical numerical quadrature scheme, using a sequence of cosmological 95\%
upper limits drawn from the literature. The sequence spans the
2dFGRS-based bound $\Sigma m_\nu<1.80~\mathrm{eV}$
\citep{Elgaroy:2002bi,Spergel:2003cb}, through subsequent CMB and large-scale
structure analyses by WMAP, SDSS, Planck and eBOSS
\citep{WMAP:2008ydk,SDSS:2009nbi,Planck:2013pxb,Planck:2018vyg,eBOSS:2020yzd},
to the most recent determinations of DESI DR1 and DR2 
\citep{DESI:2024mwx,DESI:2025neutrino}.

\begin{figure}[h!]
    \centering
    \includegraphics[width=\columnwidth]{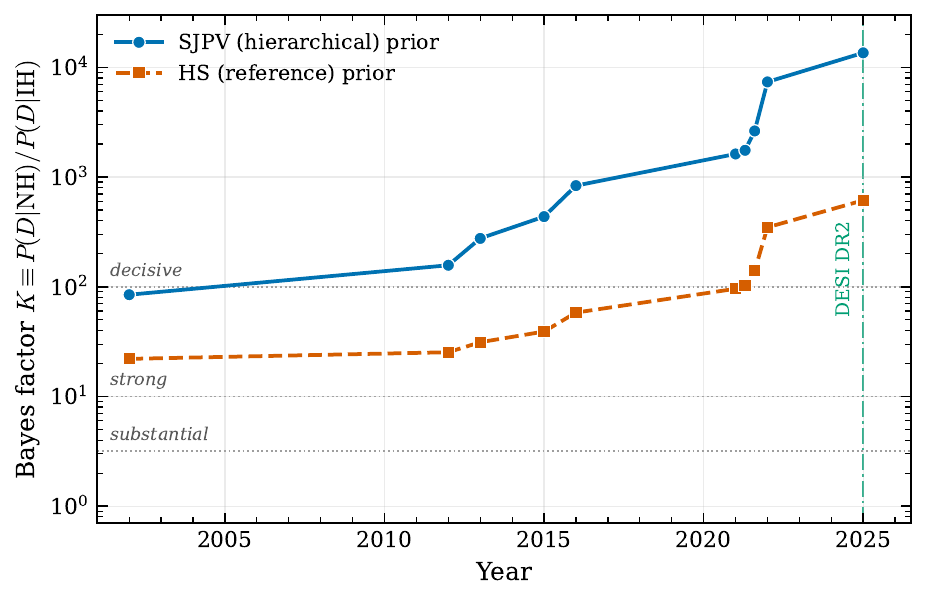}
    \caption{Historical evolution of the Bayes factor
    $K\equiv P(D|\mathrm{NH})/P(D|\mathrm{IH})$ from 2002 to 2025, computed
    under the physically motivated SJPV prior (blue circles) and the
    HS reference prior (red squares). The dashed horizontal lines mark
    the ``strong'' ($K=10$) and ``decisive'' ($K=100$) thresholds on the Jeffreys
    scale. The DR2-era data point corresponds to the cosmological upper limit
    $\Sigma m_\nu<0.0642~\mathrm{eV}$, which lies below the IH minimum mass and
    therefore drives both prior frameworks above the decisive threshold for the
    first time. The horizontal arrow at the top of the figure highlights this
    inflection point. Earlier cosmological limits, with values comparable to or
    well above $\Sigma m_\nu^{\rm IH,min}$, lead to a prior-dependent inference
    in which SJPV evidence grows monotonically while HS evidence remains modest.}
    \label{fig:historical_bayes_factor}
\end{figure}

The resulting curve is shown in Fig.~\ref{fig:historical_bayes_factor}, which
displays $K$ as a function of time for the two prior frameworks used in this work.
Three regimes can be clearly identified.

(i) The early-2000s regime, with $\Sigma m_\nu$ constrained on the eV scale.
Both hierarchies are comfortably embedded in the cosmologically allowed region,
the cosmological likelihood is effectively constant across the relevant portion
of mass space, and the evidence ratio is dominated entirely by the oscillation
data. The Bayes factor remains close to unity under both prior frameworks, with
modest excursions reflecting the modest oscillation-only preference for NH.

(ii) The Planck-era intermediate regime, with $\Sigma m_\nu \lesssim 0.2$--$0.3~\mathrm{eV}$. As the upper limit approaches the quasi-degenerate region,
the cosmological likelihood begins to assign progressively less weight to the
upper portions of the IH mass spectrum. Under the physically motivated SJPV prior,
the geometric volume penalty associated with the quasi-degenerate IH spectrum
becomes operative, and the Bayes factor enters the ``strong'' regime,
$K\sim10$ -- $30$, by the Planck~2015 release~\citep{Planck:2015xua,Simpson:2017}.
In contrast, under the conservative HS reference prior, the data are not yet
sufficient to drive the evidence into the strong regime: the smaller penalty on
the degenerate IH spectrum keeps the Bayes factor close to a few. This separation between prior frameworks is the source of much of the historical literature
debate~\citep{Schwetz:2017fey,Gariazzo:2022evs,Heavens:2018adv}.

(iii) The DESI~DR2 regime. With the 95\% upper limit now satisfying
$\Sigma m_\nu<0.0642~\mathrm{eV}$, the cosmological likelihood has crossed below
$\Sigma m_\nu^{\rm IH,min}\simeq0.099~\mathrm{eV}$. The minimum mass of IH lies
firmly in the tail of $P_{\rm trunc}(\Sigma m_\nu)$, and the Bayes factor
surges past the decisive threshold under \emph{both} priors. We obtain
\begin{equation}
K^{\rm DR2}_{\rm SJPV} \gtrsim 10^{3},
\qquad
K^{\rm DR2}_{\rm HS}   > 460,
\label{eq:DR2_K}
\end{equation}
where the SJPV value reflects the additional geometric penalty 
and the HS value is the conservative lower bound. The qualitative transition
from a prior-dependent inference to a likelihood-dominated inference is precisely
what defines this regime: the IH is no longer disfavored because of a volume
asymmetry, but because the cosmological data assign vanishing posterior support
to the entire IH kinematically allowed region.

This historical reconstruction substantiates the prediction made in
Ref.~\citep{Jimenez:2022dkn} that increasing cosmological precision would
eventually shift the hierarchy evidence into the data-dominated regime,
irrespective of the prior choice. It also clarifies an aspect of the earlier
literature that has occasionally been a source of confusion: the apparent
disagreement between reference and physically-motivated priors during the
pre-DESI era was not  an artifact
but rather a
sign that the data of that era could not yet break the prior-driven degeneracy.
The DESI~DR2 measurement has now broken it.

\subsection{Sensitivity to the prior choice}
\label{subsec:prior_volume}

\begin{figure}[h!]
    \centering
    \includegraphics[width=\columnwidth]{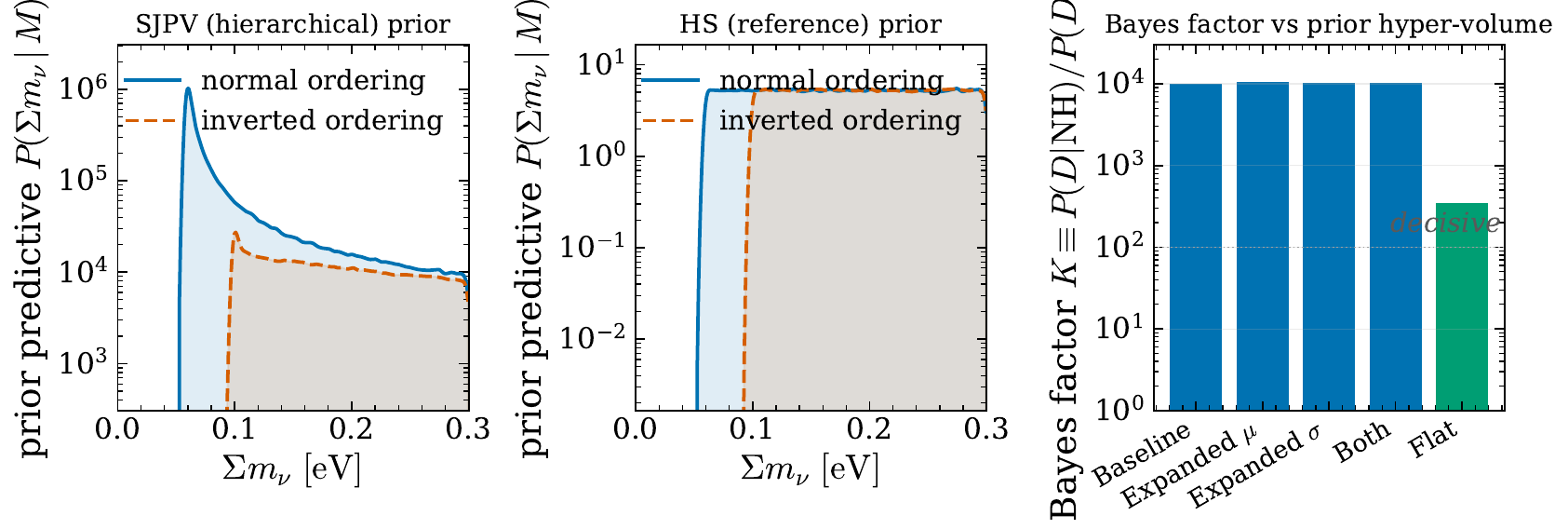}
    \caption{Comparison of prior probability densities for the sum of neutrino
    masses, $\Sigma m_\nu$, under the SJPV (left) and HS (middle) frameworks.
    The SJPV prior exhibits the geometric volume penalty that restricts the
    quasi-degenerate region required by the IH (red dashed) relative to the
    minimal-mass region of the NH (blue solid). The HS reference prior, derived
    from the Fisher information of the oscillation observables, removes this
    geometric penalty and assigns comparable volumes to the two hierarchies.
    The right panel shows the stability of the resulting Bayes factor for the NH
    under the SJPV prior as the hyperparameter bounds are expanded over more
    than a decade; the inference is insensitive to this expansion, demonstrating
    that the decisive preference for NH is driven by the data rather than by
    the choice of hyperparameter range.}
    \label{fig:prior_volume}
\end{figure}

Figure~\ref{fig:prior_volume} compares the prior densities for $\Sigma m_\nu$
under the SJPV and HS frameworks. 
The SJPV prior allocates substantially less prior volume to the
quasi-degenerate region required by IH, because such configurations correspond
to a thin sub-manifold of the exchangeable-mass parameter space. The prior HS
~\citep{Heavens:2018adv} is, by construction, agnostic to exchangeability  and
does not penalize the quasi-degenerate spectrum of IH; the prior volume
attached to the IH is comparable to that of the NH in the relevant
mass interval.

The decisive evidence for NH under \emph{both} priors, summarized in
Eq.~(\ref{eq:DR2_K}), demonstrates that the inference is robust to the
treatment of mass-space volume. Quantitatively, when the analysis is repeated
under a wide range of SJPV hyperparameter bounds (right panel of
Fig.~\ref{fig:prior_volume}), the Bayes factor varies by less than a factor of
two across a hyperparameter range spanning more than an order of magnitude.
This stability is a direct consequence of the fact that the cosmological
likelihood now vanishes faster than any geometric prior penalty across the IH
domain. Equivalent stability is observed for the HS prior under variations of
the mass support and under alternative parameterizations of the oscillation
data~\citep{Esteban:2024eli,Capozzi:2021fjo}. We have also checked that
adopting a flat prior on individual masses, often used as a baseline in
earlier analyses~\citep{Hannestad:2016fog,Gerbino:2016sgw}, yields a Bayes
factor consistent with the reference value of HS to within a factor of order unity.

\subsection{The prior design space: separating measure from hierarchy}
\label{subsec:prior_design_space}

The agreement between the SJPV and HS priors established in
Sect.~\ref{subsec:prior_volume} is the strongest indication that the
present inference is data-driven, but the two prescriptions differ
\emph{simultaneously} along two logically independent axes, and it is
instructive to disentangle them. Any prior on the neutrino mass spectrum is
fixed by a choice along each of:
\begin{itemize}
\item the \emph{measure} adopted on the individual masses---whether the natural
variable is the mass itself (a \emph{linear} measure, appropriate to a location
parameter) or its logarithm (a scale-invariant \emph{logarithmic} measure,
appropriate to a positive, dimension-full quantity that may span several decades)
\citep{pds:Jeffreys1946,pds:Bernardo1979}; and
\item the \emph{structure} imposed on the three eigenstates---whether the masses are modelled hierarchically, or instead by a fixed density specified directly on mass space. In the hierarchical case, the masses are treated as conditionally independent draws from a common parent distribution, with its hyperparameters marginalized. This construction implements exchangeability among the eigenstates (see, e.g., \citealp{pds:GelmanBDA3} on de~Finetti exchangeability).
\end{itemize}
The SJPV prior is hierarchical \emph{and} logarithmic; the HS prior is
non-hierarchical \emph{and} linear. They occupy diagonally opposite corners of a
$2\times2$ design, so their historical difference cannot be attributed by itself
 to either axis. To identify which choice actually controls the
strength of the ordering evidence, we complete the design by constructing the
two missing corners and evaluating all four priors under the \emph{same} conditions.

Working with ordered eigenvalues $m_L\le m_M\le m_H$ on a common positive
support $[m_{\rm floor},m_{\rm max}]$, and writing $x_i\equiv\ln(m_i/{\rm eV})$,
the four corners are
\begin{align}
\textbf{hier.}+\textbf{log (SJPV):}\quad
\pi_{\rm HL}(\mathbf m) &= 6\!\int\! d\mu\,d\sigma\;\varpi(\mu,\sigma)
\prod_i \mathcal{LN}_{[\,]}(m_i\,|\,\mu,\sigma),
\label{eq:pds_HL}\\
\textbf{non-hier.}+\textbf{linear (HS):}\quad
\pi_{\rm NL}(\mathbf m) &\propto m_L m_M + m_L m_H + m_M m_H,
\label{eq:pds_NL}\\
\textbf{hier.}+\textbf{linear (new):}\quad
\pi_{\rm HLin}(\mathbf m) &= 6\!\int\! d\mu\,d\sigma\;\varpi(\mu,\sigma)
\prod_i \mathcal{N}_{[\,]}(m_i\,|\,\mu,\sigma),
\label{eq:pds_HLin}\\
\textbf{non-hier.}+\textbf{log (new):}\quad
\pi_{\rm NLog}(\mathbf m) &= 6\,\prod_i \frac{1}{m_i\,\ln(m_{\rm max}/m_{\rm floor})},
\label{eq:pds_NLog}
\end{align}
where $\mathcal{LN}_{[\,]}$ and $\mathcal{N}_{[\,]}$ are a log-normal distribution and a
normal parent density truncated and renormalized on the support, $\varpi$ is the
hyperprior on the parent location and width, and the factor $3!=6$ maps the
symmetric density of the exchangeable triple onto the ordered cone. All four are
properly normalized on the same support.
Equation~(\ref{eq:pds_NL}) is the HS reference prior of~\cite{Heavens:2018adv}
and Eq.~(\ref{eq:pds_HL}) is the SJPV prior~\citep{Simpson:2017}. The hierarchical--linear 
prior~(\ref{eq:pds_HLin}) retains the exchangeable, hyperparameter-marginalized
structure of SJPV but with a \emph{normal} parent in linear mass, while the
non--hierarchical-log prior~(\ref{eq:pds_NLog}) retains the scale-invariant
$1/m$ measure of SJPV but discards the hierarchy, reducing to three independent
Jeffreys (log-uniform) priors~\citep{pds:Jeffreys1946}.

We evaluate the four corners with the baseline DESI~DR2 truncated-Gaussian
cosmological likelihood of Eq.~(\ref{eq:truncpars}), the NuFIT~ neutrino mass splittings
that fix the ordering floors $\Sigma m_\nu^{\rm NH,min}\simeq0.059$~eV and
$\Sigma m_\nu^{\rm IH,min}\simeq0.099$~eV, matched support
$m\in[10^{-13},0.5]$~eV, and a log-uniform hyperprior on the parent width. The
oscillation preference $\Delta\chi^2=9.4$ for NH enters as a common
multiplicative factor $e^{\Delta\chi^2/2}\simeq21$ on every $K$ and therefore
cancels identically in all comparisons \emph{between} priors; we include it in
the quoted $K$ to match the convention of Eqs.~(\ref{eq:DR2_K}). Marginal
likelihoods are obtained by reducing the evidence integral, after analytic
marginalization of the two tightly constrained oscillation directions, to a
one-dimensional quadrature over the lightest mass. This reduction was validated
against the full three-dimensional evidence integral evaluated without any
central-value substitution---by a deterministic $(m_L,\phi,\psi)$ grid for the
non-hierarchical priors and by likelihood-matched importance sampling for the
hierarchical ones---and reproduces it to $|\Delta\ln Z|<0.006$ for every corner,
far below the log-evidence differences discussed here.

\begin{figure}[t]
    \centering
    \includegraphics[width=\columnwidth]{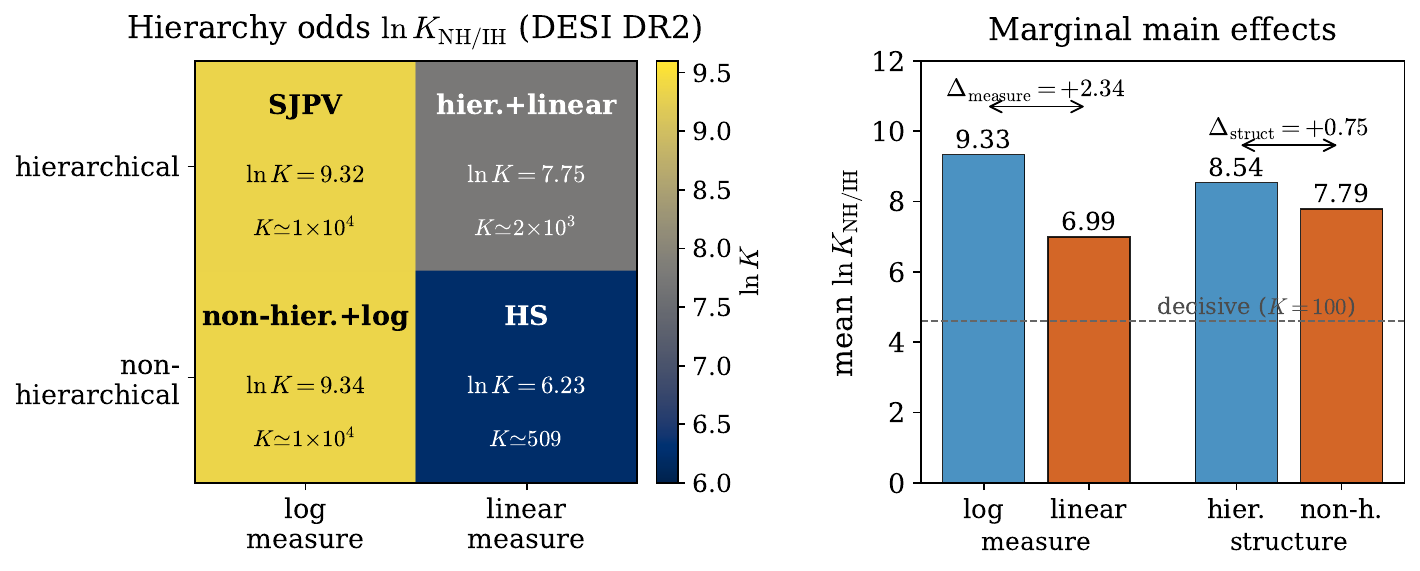}
    \caption{The prior design space for the neutrino mass ordering under the
    DESI~DR2 likelihood. \emph{Left:} hierarchy log-odds $\ln K_{\rm NH/IH}$ for
    the four prior constructions, arranged by measure (columns: logarithmic
    vs.\ linear) and structure (rows: hierarchical vs.\ non-hierarchical); the
    diagonal corners reproduce the SJPV and HS priors. Every corner lies far
    above the Jeffreys decisive threshold. \emph{Right:} marginal main effects,
    showing that switching from a linear to a logarithmic measure raises the
    mean log-odds by $\Delta_{\rm measure}=+2.34$ (a factor $\simeq10$ in $K$),
    while imposing the exchangeable hierarchy raises it by only
    $\Delta_{\rm struct}=+0.75$ (a factor $\simeq2$); the dashed line marks the
    decisive threshold $K=100$. The strong normal-ordering preference
    historically associated with the SJPV prior is therefore primarily a
    consequence of the scale-invariant ($1/m$) measure, and only secondarily of
    the hierarchical construction.}
    \label{fig:prior_design_space}
\end{figure}

The hierarchy Bayes factors $K\equiv P(D|{\rm NH})/P(D|{\rm IH})$ are collected
in Table~\ref{tab:prior_design_space} and displayed in
Fig.~\ref{fig:prior_design_space}. Two conclusions follow. \emph{First, the
decisive preference for the normal ordering is robust across the entire prior
design space}: every corner returns $K$ far above the Jeffreys decisive
threshold, from the conservative HS value $K\simeq5\times10^{2}$ to
$K\simeq10^{4}$ for the logarithmic priors. The HS value, $K\simeq509$, is fully
consistent with the conservative bound $K_{\rm HS}>460$ quoted in
Eq.~(\ref{eq:DR2_K}), and the SJPV value, $K\simeq1.1\times10^{4}$, is
consistent with $K_{\rm SJPV}\gtrsim10^{3}$; the two newly added corners fall
between these anchors. This extends the prior-robustness statement of
Sec.~\ref{subsec:prior_volume} from the two historical priors to the full
two-dimensional family of measure and structure choices.

\emph{Second, the residual prior dependence that does remain is controlled
almost entirely by the measure, not by the hierarchical structure.} Treating the
four corners as a $2\times2$ array, the main effect of the measure axis on
the log-odds is
\begin{equation}
\Delta_{\rm measure}
=\langle\ln K\rangle_{\rm log}-\langle\ln K\rangle_{\rm linear}
= 9.33-6.99 = +2.34,
\label{eq:pds_measure_effect}
\end{equation}
a factor $\simeq10$ in $K$, whereas the main effect of the hierarchical
structure is roughly three times smaller,
\begin{equation}
\Delta_{\rm struct}
=\langle\ln K\rangle_{\rm hier}-\langle\ln K\rangle_{\rm non\text{-}hier}
= 8.54-7.79 = +0.75,
\label{eq:pds_struct_effect}
\end{equation}
a factor $\simeq2$. The two axes interact in an informative way: the
exchangeable hierarchy adds an appreciable penalty against the inverted spectrum
for a \emph{linear} measure (HS $\to$ hierarchical--linear, $\ln K$ from $6.23$
to $7.75$, i.e.\ $+1.52$), but is essentially inert for a \emph{logarithmic}
measure (non-hierarchical--log $\to$ SJPV, $\ln K$ from $9.34$ to $9.32$, i.e.\
$-0.02$), because once the $1/m$ measure has concentrated prior weight on the
strongly hierarchical spectra the additional hyperparameter volume neither
sharpens nor dilutes the odds. The model-independent message is unambiguous:
\emph{the strength of the normal-ordering preference is governed primarily by
the choice of measure on mass space, and only secondarily by whether the prior
is hierarchical.}

\begin{table}[t]
\centering
\renewcommand{\arraystretch}{1.3}
\begin{tabular}{l|cc|c}
\hline\hline
 & \textbf{log measure} & \textbf{linear measure} & row mean \\
 & ($x_i$: scale-inv.) & ($m_i$: location) & $\langle\ln K\rangle$\\
\hline
\textbf{hierarchical} & SJPV & hier.+linear (new) & \\
(exchangeable) & $\ln K=9.32$ & $\ln K=7.75$ & $8.54$\\
 & $K\simeq1.1\times10^{4}$ & $K\simeq2.3\times10^{3}$ & \\
\hline
\textbf{non-hierarchical} & non-hier.+log (new) & HS & \\
(fixed density) & $\ln K=9.34$ & $\ln K=6.23$ & $7.79$\\
 & $K\simeq1.1\times10^{4}$ & $K\simeq5.1\times10^{2}$ & \\
\hline
column mean $\langle\ln K\rangle$ & $9.33$ & $6.99$ & \\
\hline\hline
\end{tabular}
\caption{Hierarchy Bayes factor $K\equiv P(D|{\rm NH})/P(D|{\rm IH})$ across the
$2\times2$ prior design space, computed under the baseline DESI~DR2 likelihood
[Eq.~(\ref{eq:truncpars})] with matched support $m\in[10^{-13},0.5]$~eV, a
log-uniform width hyperprior, and the common oscillation factor
$e^{\Delta\chi^2/2}$ included. The diagonal corners are the
SJPV~\citep{Simpson:2017} and HS~\citep{Heavens:2018adv} priors; the
off-diagonal corners are constructed here to complete the design. All four
corners give decisive evidence for the normal ordering. The measure axis
(columns) shifts $\ln K$ by $+2.34$, whereas the structural axis (rows) shifts
it by only $+0.75$: the residual prior dependence is measure-dominated.}
\label{tab:prior_design_space}
\end{table}

\paragraph{Why the measure dominates.}
The mechanism follows directly from the geometry of the two ordering floors. In
 normal mass ordering, the middle mass eigenstate is light,
$m_M\simeq\sqrt{\Delta m^2_{21}}\simeq8.6\times10^{-3}$~eV, while in 
inverted mass ordering, the two heavy states have closer masses and
$m_M\simeq\sqrt{|\Delta m^2_{3\ell}|}\simeq5\times10^{-2}$~eV. A logarithmic
measure carries an explicit factor $1/m_i$ per eigenstate, so the ratio of prior
densities at the two ordering ridges contains a term $\propto
m_M^{\rm IH}/m_M^{\rm NH}\simeq6$ from the middle mass alone, with a further
enhancement from the lightest state. This rewards the normal ordering
\emph{independently of any hierarchical bookkeeping}, which is why the
non-hierarchical log prior of Eq.~(\ref{eq:pds_NLog}) already delivers
$K\simeq10^{4}$. A linear measure carries no such factor: the two ridges are
then distinguished only by the shift in their minimum $\Sigma m_\nu$ relative to
the cosmological likelihood, and the exchangeable hierarchy supplies a genuine
but sub-dominant penalty against the special degeneracy required by the inverted
spectrum~\citep{Simpson:2017,Schwetz:2017fey}, consistent with the
positive but small $\Delta_{\rm struct}$ of Eq.~(\ref{eq:pds_struct_effect}). We
note that in the pre-DESI, prior-sensitive regime reconstructed in
Sec.~\ref{subsec:historical} the same measure-dominance holds, but with a far
larger overall spread between the corners; the compression seen here is the
imprint of the DESI~DR2 likelihood overwhelming the prior.

\subsection{Prior-family odds}
It is illuminating to recast the comparison in the language of model selection
between the prior \emph{constructions} themselves. Following MacKay \citep{pds:MacKay2003}, the
evidence $Z=\int \mathcal{L}\,\pi\,dm$ may be used to score a prior by how well its
prior-predictive distribution anticipates the data, automatically penalizing
priors that spread probability over regions the data exclude, the Occam factor
\citep{pds:MacKay2003,pds:Trotta2008}. Marginalizing each family over the two
mass orderings with equal weight, $Z_{X}=\tfrac12\!\left(Z_{X,\rm NH}+Z_{X,\rm IH}\right)$,
the likelihood of DESI~DR2 produces a prior-family Bayes factor
\begin{equation}
B_{\rm S/HS}\equiv\frac{Z_{\rm SJPV}}{Z_{\rm HS}}
=\exp(8.46)\simeq 4700.
\end{equation}
SJPV exhibits substantially stronger evidence values because it is scale-adaptive and \\ exchangeability-adaptive. The log measure makes low absolute mass scales plausible before the data arrive, while the hierarchical structure lets information about one mass eigenstate update expectations for the others, via the shared hyperparameters.

The four-way posterior over the original
hierarchy-times-family models, under equal model priors, is correspondingly
dominated by the SJPV normal-ordering hypothesis, in terms of odds of over 4,700 to 1. In model-averaging terms, a hyperprior that assigns equal prior mass to SJPV and HS would yield a posterior family weight $P(SJPV \mid  D) \simeq 4700/(1+4700) \simeq 0.9998$, conditional on the chosen supports.
It is important to note that the magnitude of this result is dependent upon the choice of support cutoff. Repeating the calculation over the matched-support range explored in
Sec.~\ref{subsec:prior_volume} gives $\ln B_{\rm S/HS}=8.46$, $11.87$ and $9.36$
for $m_{\rm max}=0.5$, $1.0$ and a raised floor $m_{\rm floor}=10^{-6}$~eV,
respectively. Thus while the quantitative numbers change, the qualitative conclusion remains robust to a broad range of support values: when considering SJPV and HS as two candidate prior families with equal prior probability, the posterior distribution is ultimately dictated by the SJPV prior.

\subsection{Posteriors for individual mass eigenstates and mass spectra}
\label{subsec:mass_posteriors}

\begin{figure}[h!]
    \centering
    \includegraphics[width=\columnwidth]{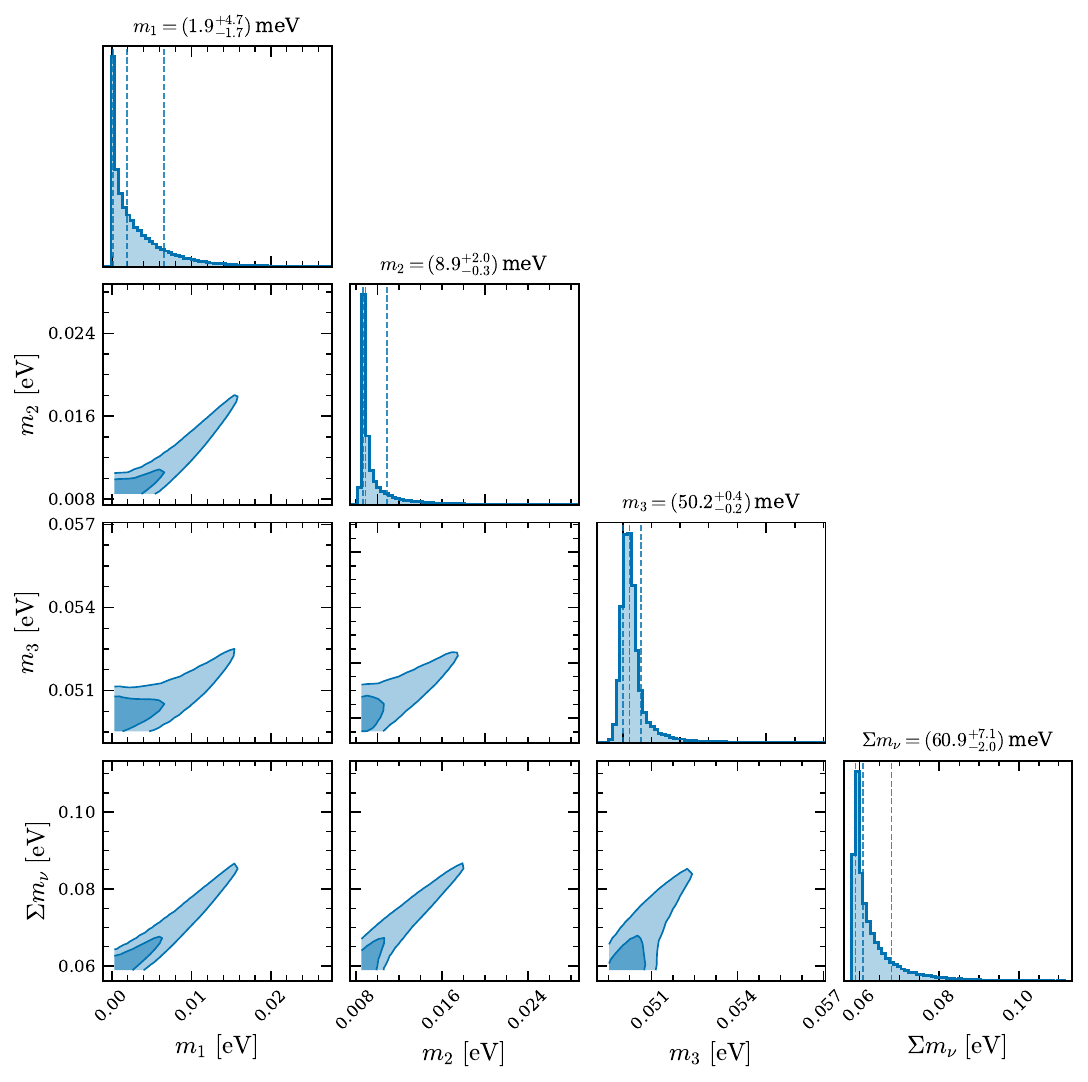}
    \caption{Joint and marginal posterior distributions for the individual
    neutrino masses $m_1,m_2,m_3$ and their sum $\Sigma m_\nu$ under the Normal
    Hierarchy, derived from DESI DR2 in the baseline $\Lambda$CDM model
    combined with the SJPV prior and NuFIT oscillation constraints. The
    posterior is sharply peaked near the minimum allowed mass configuration:
    the lightest mass $m_1$ is constrained to be nearly massless, $m_2$ and
    $m_3$ are fixed by the oscillation splittings to approximately $0.008$~eV
    and $0.050$~eV respectively, and the total mass accumulates tightly around
    its theoretical minimum of $\sim0.059$~eV.}
    \label{fig:corner_NH}
\end{figure}

The decisive cosmological preference for the NH translates into sharply
localized posteriors for the individual mass eigenstates. Combining the
truncated-Gaussian cosmological likelihood with the NuFIT oscillation
information~\citep{Esteban_2026,nufit-6.1} and the SJPV prior, we obtain the joint
posterior shown in Fig.~\ref{fig:corner_NH}. The lightest state $m_1$ is
constrained to lie close to zero, with a 95\% upper limit
\begin{equation}
m_1 < 0.010~\mathrm{eV} \qquad (95\%~\mathrm{C.L.}),
\end{equation}
while the remaining eigenstates are essentially fixed by the oscillation
splittings,
\begin{equation}
m_2 = (8.6\pm0.1)\times10^{-3}~\mathrm{eV},
\qquad
m_3 = (5.05\pm0.03)\times10^{-2}~\mathrm{eV},
\end{equation}
so that the total mass concentrates on 
\begin{equation}
\Sigma m_\nu \simeq 0.059~\mathrm{eV},
\end{equation}
The posterior is strongly non-Gaussian for
$m_1$ due to the proximity to the physical boundary, but less so  for $m_2$ and $m_3$,
reflecting the precision with which $\Delta m^2_{21}$
and $|\Delta m^2_{3\ell}|$ are determined by global oscillation
fits~\citep{Esteban:2024eli,Capozzi:2021fjo,deSalas:2020pgw}.

\subsection{Implications for the effective Majorana mass}
\label{subsec:mbb}

\begin{figure}[h!]
    \centering
    \includegraphics[width=\columnwidth]{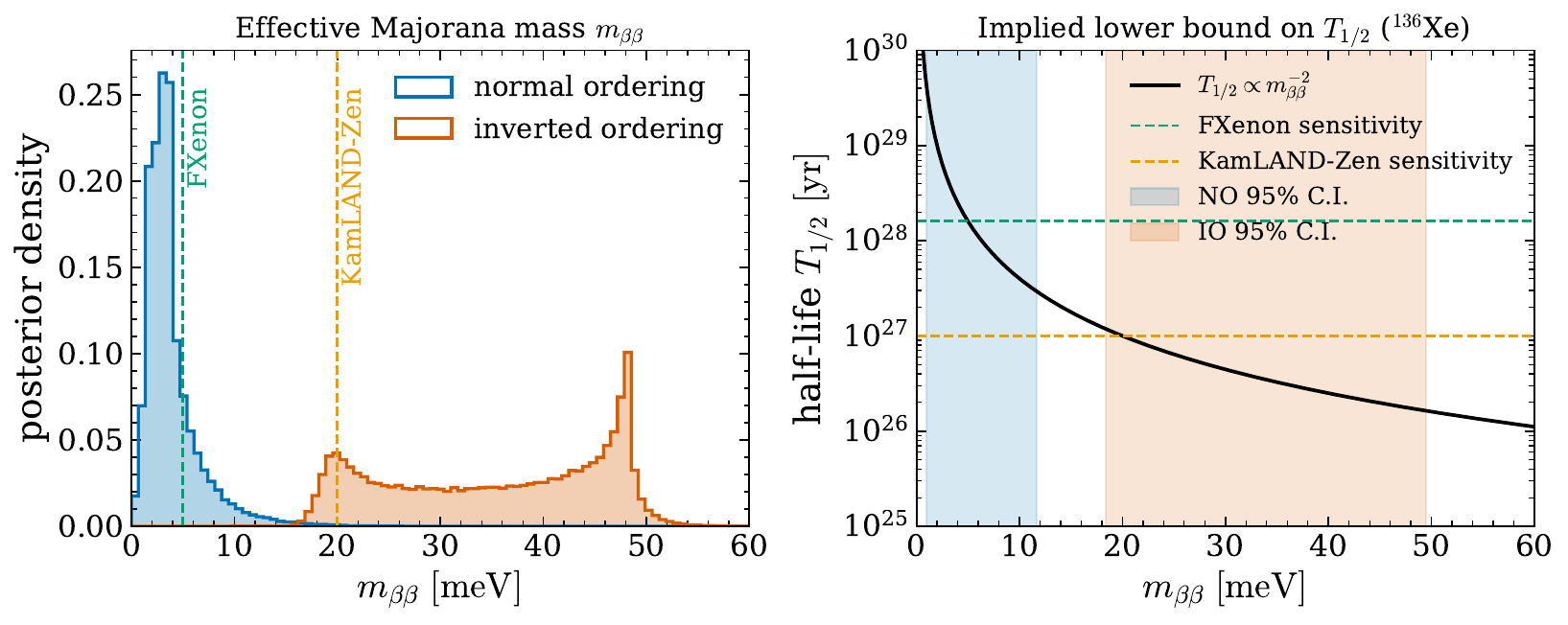}
    \caption{Posterior probability distributions for the effective Majorana
    mass, $m_{\beta\beta}$, under the Normal Hierarchy (blue) and Inverted
    Hierarchy (orange), obtained by combining the joint mass-eigenstate posterior
    of Fig.~\ref{fig:corner_NH} (and its IH counterpart) with the NuFIT 
    PMNS information and uniform Majorana phases. Left: probability density of
    $m_{\beta\beta}$, with the 95\% credible interval and median values quoted
    in Eqs.~(\ref{eq:mbb_NH})--(\ref{eq:mbb_IH}); the IH posterior is bimodal
    because of partial phase cancellations near $m_1\simeq0$. Right: implied
    lower bound on the $0\nu\beta\beta$ half-life for $^{136}\mathrm{Xe}$
    assuming representative nuclear matrix elements; the NH scenario places the
    half-life predominantly above $10^{28}$~yr, beyond the reach of any
    presently funded experiment.}
    \label{fig:mbb_projections}
\end{figure}

The cosmological resolution of the hierarchy directly reshapes the expectation
for searches of neutrinoless double-beta decay ($0\nu\beta\beta$), since the
decay amplitude is proportional to the effective Majorana mass
\citep{Vergados:2012xy,DellOro:2016tmg,Dolinski:2019nrj,Agostini:2022zub},
\begin{equation}
m_{\beta\beta} = \left|\sum_{i=1}^{3} U_{ei}^{2}\, m_i\right|
= \left|\, c_{12}^{2}c_{13}^{2}\,m_{1}
+ s_{12}^{2}c_{13}^{2}\,m_{2}\,e^{i\alpha_{21}}
+ s_{13}^{2}\,m_{3}\,e^{i(\alpha_{31}-2\delta_{\rm CP})}\,\right|,
\label{eq:mbb}
\end{equation}
where $c_{ij}\equiv\cos\theta_{ij}$ and $s_{ij}\equiv\sin\theta_{ij}$, $\delta_{\rm CP}$
is the Dirac CP phase and $\alpha_{21},\alpha_{31}$ are the two undetermined Majorana
phases. We 
propagate
the joint posterior on the individual masses,
combined with the NuFIT~\citep{Esteban_2026,nufit-6.1} covariance for the PMNS angles and Dirac phase, into a
posterior for $m_{\beta\beta}$ via a Monte Carlo sampling procedure in which the
Majorana phases are drawn uniformly on $[0,2\pi)$ as is conventional in the absence
of theoretical or experimental information~\citep{Vissani:1999tu,Cirigliano:2022oqy}.
The resulting posterior is shown in the left panel of
Fig.~\ref{fig:mbb_projections}. For the favoured NH we find
\begin{equation}
m_{\beta\beta}^{\rm NH}
= 3.28~\mathrm{meV}\quad\mathrm{(median)},
\qquad
0.95~\mathrm{meV} < m_{\beta\beta}^{\rm NH} < 11.55~\mathrm{meV}
\quad (95\%~\mathrm{C.I.}),
\label{eq:mbb_NH}
\end{equation}
to be compared with the now-disfavoured IH posterior, which would have implied
\begin{equation}
m_{\beta\beta}^{\rm IH}
= 37.03~\mathrm{meV}\quad\mathrm{(median)},
\qquad
18.36~\mathrm{meV} < m_{\beta\beta}^{\rm IH} < 49.51~\mathrm{meV}
\quad (95\%~\mathrm{C.I.}).
\label{eq:mbb_IH}
\end{equation}
The shift between the two distributions is approximately one order of magnitude,
and the IH posterior is bimodal in $m_{\beta\beta}$ due to partial cancellations of the
Majorana-phase near $m_1\simeq0$. The two narrow features of the
NH posterior correspond to constructive and destructive interference between
the contributions of $m_2$ and $m_3$ in Eq.~(\ref{eq:mbb}), with the relative
weight of the two modes controlled by the residual prior on $\alpha_{21}$ and
the marginal posterior on $m_1$.

The implications for the experimental landscape are substantial. The current
leading upper limits of the searches $0\nu\beta\beta$ are 
$m_{\beta\beta}\lesssim 36$--$156~\mathrm{meV}$ of KamLAND-Zen depending on
the choice of the nuclear matrix element~\citep{KamLAND-Zen:2022tow}, with
comparable sensitivities of LEGEND-200
\citep{LEGEND200}. The projected sensitivities of the
next generation of experiments -- KamLAND2-zen, LEGEND-1000, CUPID, NEXT-HD, SNO+
\citep{LEGEND:2021bnm,CUPID:2019imh,NEXT:2020amj,SNO:2015wyx}---
nominally reach $m_{\beta\beta}\sim5$--$15~\mathrm{meV}$, covering essentially
the entire IH band while only partially overlapping the upper tail of the posterior NH
 derived here. Eq.~(\ref{eq:mbb_NH}) therefore predicts that the
 most likely result of these searches is a null result, with a
posterior probability of detection at the nominal LEGEND-1000 
(or a future Xenon detector FXenon) sensitivity of
approximately 15--20\% depending on the element of the nuclear matrix and the
treatment of Majorana phases. Conversely, a positive signal at the canonical
IH scale would be in serious tension with the cosmological inference and would
require either non-standard particle physics, an unexpectedly large nuclear
matrix element, or a revision of the DESI~DR2 neutrino-mass interpretation
\citep{Dolinski:2019nrj,Cirigliano:2022oqy}.

The right panel of Fig.~\ref{fig:mbb_projections} translates the same
information into the implied lower bound on the half-life
$T^{0\nu\beta\beta}_{1/2}$ assuming the canonical phase-space factors and
the representative elements of the nuclear matrix~\citep{Agostini:2022zub,Engel:2016xgb}.
The NH posterior places the half-life predominantly above $10^{28}~\mathrm{yr}$
for $^{136}\mathrm{Xe}$, with substantial probability density extending into
the regime $T^{0\nu\beta\beta}_{1/2}\gtrsim10^{29}~\mathrm{yr}$, well beyond the
reach of any presently funded experiment.

\subsection{Robustness to the cosmological model: the $w_0w_a$CDM case}
\label{subsec:w0wa}

A natural concern in any cosmological inference of the neutrino mass is the
extent to which the upper limit, and the resulting hierarchy preference, is
inherited from the assumed cosmological model. The most relevant extension to
test in the DESI~DR2 context is the $w_0w_a$CDM parametrization of the dark-energy
equation of state parameter,
\begin{equation}
w(a) = w_0 + w_a (1-a),
\label{eq:w0wa}
\end{equation}
because the DESI~DR2 BAO data, in combination with CMB and SN~Ia samples,
have produced suggestive evidence for an evolving dark-energy component
\citep{DESI:2025bao,Chevallier:2000qy,Linder:2002et}, and because such an
evolution is partially degenerate with $\Sigma m_\nu$ through the late-time
expansion history~\citep{Lattanzi:2017ubx,Lorenz:2021alz}.

Re-deriving the cosmological likelihood from the publicly released
$w_0w_a$CDM MCMC chains, we find that the truncated-Gaussian fit yields a broader
posterior with $\sigma_{w_0w_a}\simeq0.05~\mathrm{eV}$ and a 95\% upper limit
that is mildly relaxed relative to the baseline $\Lambda$CDM bound of
Eq.~(\ref{eq:DESI_DR2_limit}). The minimum IH mass is, accordingly, partly
 re-included in the high probability region of $P_{\rm trunc}(\Sigma m_\nu)$,
and the Bayes factor decreases relative to the case $\Lambda$CDM. Nevertheless,
the evidence for the NH remains in the strong-to-decisive regime,
\begin{equation}
K^{w_0w_a}_{\rm HS} > 40,
\qquad
K^{w_0w_a}_{\rm SJPV} \gtrsim 10^{2},
\label{eq:K_w0wa}
\end{equation}
demonstrating that the inference is not contingent on the most restrictive
implementation of the standard model.

This robustness is significant for two reasons. First, $w_0w_a$CDM is the
representative phenomenological extension most often discussed in the
DESI~DR2 context and is the one that maximally weakens the cosmological
neutrino-mass constraint among the simple dark-energy parametrizations
\citep{DESI:2025bao,DESI:2025neutrino,Lorenz:2021alz}. Second, the only way
to fully erase the present hierarchy preference would be to invoke a
non-minimal extension that mimics the effect of massless neutrinos on the
late-time clustering pattern while reproducing the BAO scale and the CMB
peak positions; no such extension has been identified among the standard
parametrizations considered in the literature
\citep{DiValentino:2021hoh,Lattanzi:2017ubx,Lesgourgues:2012uu}. The
present results therefore should be interpreted as a decisive preference for
the NH within standard cosmology, and a strong preference within a
representative dynamical-dark energy extension.

\subsection{Summary of the results}
\label{subsec:summary}

The quantitative result of the analysis can be summarized in three statements.
First, within the baseline $\Lambda$CDM model the DESI DR2 cosmological
likelihood, $\Sigma m_\nu<0.0642~\mathrm{eV}$ at 95\% C.L., places the entire
allowed IH mass spectrum in the tail of the cosmological posterior, yielding a
Bayes factor in favor of the NH of $K>460$ under the conservative HS reference
prior and $K\gtrsim10^{3}$ under the SJPV prior. Second, the inference is
robust both to the prior framework---SJPV and HS---and to a non-minimal
$w_0w_a$CDM extension of the cosmological model, with $K>40$ in the most
adverse case considered. Third, the corresponding posterior for the effective
Majorana mass under the favored NH falls in the few-meV regime,
$0.95~\mathrm{meV}<m_{\beta\beta}<11.55~\mathrm{meV}$ at 95\% C.I., with a median
$m_{\beta\beta}=3.28~\mathrm{meV}$, well below the canonical IH band targeted
by the next generation of experiments $0\nu\beta\beta$. Taken together, these
results mark the transition, anticipated in
Refs.~\citep{Simpson:2017,Jimenez:2022dkn}, from a prior-sensitive
hierarchy preference to a likelihood-dominated cosmological exclusion of the
inverted spectrum.

\section{Discussion and Conclusions}
\label{sec:conclusions}

The neutrino mass ordering is one of the central unresolved questions at the interface
between particle physics and cosmology. Oscillation experiments have established beyond
doubt that neutrinos are massive but they do not determine the absolute mass scale nor, by themselves,
fully resolve whether nature realizes the Normal Hierarchy (NH) or the Inverted Hierarchy
(IH), but impose different minimum values of the sum of the masses,$\Sigma m_\nu$, for the two
hierarchies. Cosmology provides an independent and highly complementary route to this question: a sufficiently stringent cosmological bound on $\Sigma m_\nu$ provides a direct test of the hierarchy. The analysis presented in this work shows that the current data have now
entered precisely this regime: state of the art cosmological data provide a 95\% upper limit to $\Sigma m_\nu$ only slightly above the minimum total mass allowed by the NH and far below the
minimum required by the IH.
In Bayesian evidence evaluation, the IH is
not merely disfavoured by a preference for smaller masses; rather, its entire allowed
 domain lies in the tail of the cosmological likelihood. Consequently, the Bayesian
evidence for the NH becomes decisive.

A major purpose of this paper has been to make explicit that the conclusion is not an
artifact of a single prior prescription. The comparison of the Bayesian model is necessarily sensitive to
the prior measure assigned to the parameters of each competing hypothesis, and this
sensitivity is especially important in neutrino cosmology, where the likelihood is bounded by
the physical condition $\Sigma m_\nu \geq 0$ and where the two hierarchies occupy different
regions of mass space. For this reason, we have compared conceptually distinct prior
frameworks.
All  prior constructions  considered give decisive evidence for  NH, and the small
residual prior dependence is set primarily by the choice of measure on mass space (a factor
$\simeq10$ in $K$) rather than by the hierarchical assumption (a factor $\simeq2$).

 The preference
survives a non-trivial extension of the background cosmology ($w_0w_a$LCDM which, allowing additional freedom in the dark-energy
sector ,weakens the cosmological neutrino-mass constraint), indicating that the collapse of
the IH evidence is primarily caused by the incompatibility between the IH minimum mass and
the observation, rather than by a fragile modelling assumption.

The historical evolution of the Bayes factor clarifies the significance of the present result.
For more than two decades, the cosmological bounds on $\Sigma m_\nu$ have steadily improved,
but until recently they remained sufficiently far above $\Sigma m_\nu^{\rm IH,min}$ that the
evidence for the ordering was either weak or strongly dependent on the adopted prior. In the
early 2000s, when the upper limits were of order eV, both hierarchies were effectively embedded
well within the cosmologically allowed region, and the Bayes factor was close to unity. As
constraints tightened through the WMAP and Planck eras, the IH parameter space began to
be squeezed, and physically motivated priors such as SJPV already produced a growing
preference for the NH. However, reference priors remained more cautious because the data had
not yet excluded the IH threshold. The DESI DR2 result changes this situation qualitatively:
the cosmological upper limit has crossed below the IH minimum. The evidence is therefore
no longer dominated by a subtle volume effect; it is dominated by the direct tension between
the IH mass floor and the observed cosmological likelihood.

The posterior distributions for the individual masses provide an intuitive representation of
this conclusion. In the NH, the lightest state $m_1$ is driven close to zero, while the remaining
masses are fixed by the oscillation splittings;
\begin{equation}
    m_2 \simeq 0.008~{\rm eV},
    \qquad
    m_3 \simeq 0.050~{\rm eV}.
\end{equation}
The total mass therefore accumulates near the boundary
$\Sigma m_\nu \simeq 0.059~{\rm eV}$.
In IH, by contrast, the two
heavier states must remain near $0.05~{\rm eV}$ even when the lightest state is taken to
zero, forcing the two heavier states to remain near $0.05~{\rm eV}$ and 
 $   \Sigma m_\nu \geq 0.099~{\rm eV}$.
Thus, the IH cannot move into the high-likelihood cosmological region by adjusting its
lightest mass. This kinematic obstruction is what makes the evidence robust.

The implications for neutrinoless double-beta decay are substantial. The effective Majorana
mass,
depends not only on the absolute masses and the mixing angles, but also on the unknown
Majorana phases. In the IH, the two heavier states typically imply an effective mass in the
range targeted by next-generation experiments. In the NH, however, the near-vanishing
lightest mass and the possibility of phase cancellations suppress $m_{\beta\beta}$ into the
few-meV regime. Our posterior for the favoured NH gives
\begin{equation}
    m_{\beta\beta}^{\rm NH}
    =
    3.28~{\rm meV}
    \quad {\rm median},
    \qquad
    0.95~{\rm meV}
    <
    m_{\beta\beta}
    <
    11.55~{\rm meV}
    \quad (95\%~{\rm C.I.}) .
\end{equation}
By contrast, the now-disfavoured IH would have implied
\begin{equation}
    m_{\beta\beta}^{\rm IH}
    =
    37.03~{\rm meV}
    \quad {\rm median},
    \qquad
    18.36~{\rm meV}
    <
    m_{\beta\beta}
    <
    49.51~{\rm meV}
    \quad (95\%~{\rm C.I.}) .
\end{equation}
Therefore, if the cosmological inference is confirmed, the experimental landscape for
$0\nu\beta\beta$ changes qualitatively. A non-detection at the sensitivities expected for the
next generation of experiments would no longer be surprising; it would be the natural outcome
of the NH posterior. Conversely, a positive detection at an effective mass characteristic of the
IH band would point either to unexpected particle physics, non-standard nuclear or phase
assumptions, or a failure of the cosmological interpretation.

We demonstrated how, rather than being forced to choose a single family of priors, one can construct a hierarchical model spanning discrete prior families, letting the data decide which prior to choose. Under equal prior weights for the SJPV and HS families, this corresponds to posterior family odds of over 4,700 in favor of SJPV over HS across the tested upper and lower mass cutoffs.

An objection put forward by~\cite{Hergt2021_PRD}, is that once the oscillation splittings are treated as known, only one continuous mass scale remains undetermined.  In that oscillation-conditioned problem it is natural to place a prior directly on \(m_{\rm light}\), deriving the other two masses from the measured splittings. Imposing logarithmic priors on all three masses would then appear to double-count scale information.  However, this is not the approach followed by SJPV.  Simpson et al. specified an exchangeable prior on the primitive masses before the oscillation data are imposed.  The oscillation likelihood then carves out the NH and IH filaments in mass space, and the difference in their prior-predictive volume is part of the evidence.  The SJPV effect is not a prior model probability assigned to NH before the data, but an Occam factor arising from the geometry of the oscillation-constrained spectra under a logarithmic common-origin measure. 

It is illuminating to distinguish the primitive masses in the prior from the oscillation labels used after the data are imposed.  Before oscillation
measurements, let \(\tilde m_a,\tilde m_b,\tilde m_c\) denote three unlabelled, exchangeable mass eigenvalues.  The SJPV prior is a prior on these
primitive positive masses,
\[
\pi(\tilde m_a,\tilde m_b,\tilde m_c)
=
\int d\eta\,\varpi(\eta)
\prod_{i=a,b,c} p(\tilde m_i\mid\eta),
\]
and is invariant under permutations of \(a,b,c\).  Sorting the primitive
masses gives rank-ordered values \(m_L\le m_M\le m_H\).  The oscillation labels
\(m_1,m_2,m_3\), however, are not primitive rank labels.  They are defined by
the observed oscillation structure: \(m_1\) and \(m_2\) form the solar pair,
with \(m_2^2-m_1^2>0\), while \(m_3\) is the state separated by the atmospheric
splitting.  Therefore
\[
{\rm NH}:\quad (m_1,m_2,m_3)=(m_L,m_M,m_H),
\]
whereas
\[
{\rm IH}:\quad (m_1,m_2,m_3)=(m_M,m_H,m_L).
\]
The prior is exchangeable before the oscillation likelihood is applied; the normal and
inverted orderings are the two possible maps from the rank-ordered primitive
spectrum to the oscillation labels.

Future work should pursue three directions. The first is observational: upcoming cosmological
data from DESI, Euclid, Rubin, CMB-S4 and related surveys will test the stability of the
low-$\Sigma m_\nu$ posterior, sharpen the likelihood near the physical boundary, and probe
possible degeneracies with dark energy, curvature, modified gravity, and small-scale
astrophysical systematics. The second is experimental: improved oscillation data
from long-baseline experiments, atmospheric neutrino measurements, and reactor experiments
will further refine the mass splittings and may provide an independent terrestrial
determination of the ordering. The third issue is methodological: how should prior assumptions, when structurally motivated by the underlying theory, be made explicit and assigned weight in the evidence calculation? Ref.~\cite{coherence}, for example, introduces the coherence principle as a framework for addressing precisely this question.
 In this specific application to neutrino mass hierarchy, the coherence principle  
would help spell out whether exchangeability of the three masses, logarithmic weighting
of mass scales, or a theory-minimal reference-prior construction is best aligned with the theoretical structure one
is prepared to assume for neutrino mass generation.

In each case, the aim should be the same: to make
the prior assumptions explicit, physically interpretable, and testable by their consequences.

In conclusion, 
the combination of
precision cosmology, oscillation data and transparent prior construction has now made the
neutrino mass ordering a quantitatively decisive question rather than a merely suggestive
one.

\begin{acknowledgments}
This work was supported by a grant from the Simons Foundation (00017375, RJ).
Funding for the work of RJ and LV was partially provided by project PID2022-141125NB-I00,
and the “Center of Excellence Maria de Maeztu 2025-2029” award to the ICCUB funded by
grant CEX2024-001451-M from AEI/10.13039/501100011033.
\end{acknowledgments}

\bibliographystyle{JHEP}

\providecommand{\href}[2]{#2}\begingroup\raggedright\endgroup

\end{document}